\documentclass[twocolumn]{aastex63}

\usepackage{amsmath}
\usepackage{bm}
\usepackage{multirow}

\graphicspath{{./}{Figures/}}
\shorttitle{Asymmetric nuclear matter in relativistic mean-field models}
\shortauthors{Miyatsu, Cheoun, and Saito}

\begin{document}

\title{Asymmetric nuclear matter in relativistic mean-field models with isoscalar- and isovector-meson mixing}

\correspondingauthor{Tsuyoshi Miyatsu}
\email{tsuyoshi.miyatsu@ssu.ac.kr}

\author[0000-0001-9186-8793]{Tsuyoshi Miyatsu}
\affiliation{Department of Physics and OMEG Institute, Soongsil University, Seoul 06978, Republic of Korea}

\author[0000-0001-7810-5134]{Myung-Ki Cheoun}
\affiliation{Department of Physics and OMEG Institute, Soongsil University, Seoul 06978, Republic of Korea}

\author[0000-0002-8563-9262]{Koichi Saito}
\affiliation{Department of Physics, Faculty of Science and Technology, Tokyo University of Science, Noda 278-8510, Japan}

\begin{abstract}
  Using the relativistic mean-field model with nonlinear couplings between the isoscalar and isovector mesons, we study the properties of isospin-asymmetric nuclear matter.
  Not only the vector mixing, $\omega_{\mu}\omega^{\mu}\bm{\rho}_{\nu}\bm{\rho}^{\nu}$, but also the quartic interaction due to the scalar mesons, $\sigma^{2}\bm{\delta}^{2}$, is taken into account to investigate the density dependence of nuclear symmetry energy, $E_{\rm sym}$, and the neutron-star properties.
  It is found that the $\delta$ meson increases $E_{\rm sym}$ at high densities, whereas the $\sigma$-$\delta$ mixing makes $E_{\rm sym}$ soft above the saturation density.
  Furthermore, the $\delta$ meson and its mixing have a large influence on the radius and tidal deformability of a neutron star.
  In particular, the $\sigma$-$\delta$ mixing reduces the neutron-star radius, and, thus, the present calculation can simultaneously reproduce the dimensionless tidal deformabilities of a canonical $1.4M_{\odot}$ neutron star observed from the binary neutron star merger, GW170817, and from the compact binary coalescence, GW190814.
\end{abstract}


\section{Introduction} \label{sec:introduction}

Relativistic mean-field (RMF) calculations have been widely adopted to provide a realistic description of the bulk properties of finite nuclei and nuclear matter \citep{Chin:1974sa,Walecka:1974qa}.
They are still essential methods for understanding high-energy phenomena and/or dense nuclear matter because it is necessary to treat the nuclear equation of state (EoS) relativistically \citep{Glendenning:1991es,Li:2008gp}.

Based on the one-boson exchange (OBE) potential for nuclear interactions \citep{Machleidt:1987hj,Machleidt:1989tm}, the original RMF model has been constructed by the exchange of isoscalar, Lorentz-scalar ($\sigma$) and Lorentz-vector ($\omega^{\mu}$) mesons between nucleons \citep{Serot:1984ey}.
The nonlinear self-coupling of $\sigma$ and $\omega$ mesons has been also introduced to reproduce a reasonable nuclear incompressibility and/or properties of unstable nuclei \citep{Boguta:1977xi,Sugahara:1993wz,Lalazissis:1996rd}.
In addition, the isovector, Lorentz-vector ($\bm{\rho}^{\mu}$) meson and its nonlinear couplings, e.g., $(\bm{\rho}_{\mu}\bm{\rho}^{\mu})^{2}$, $\sigma^{2}\bm{\rho}_{\mu}\bm{\rho}^{\mu}$, and $\omega_{\mu}\omega^{\mu}\bm{\rho}_{\nu}\bm{\rho}^{\nu}$, have been considered to describe a neutron skin thickness of heavy nuclei and characteristics of isospin-asymmetric nuclear matter \citep{Mueller:1996pm,Horowitz:2000xj,Horowitz:2001ya}.
At present, many kinds of the RMF models with nonlinear couplings are used to study compact-star physics as well as nuclear physics \citep{Dutra:2014qga,Choi:2021odz,Kumar:2021vdr}.

Owing to the experimental analyses of heavy-ion collisions, the nuclear symmetry energy, $E_{\rm sym}$, and its slope parameter, $L$, turn out to play very important roles in determining the nuclear EoS for isospin-asymmetric matter \citep{Typel:2001lcw,Danielewicz:2002pu,Lattimer:2004pg,Steiner:2004fi,Tsang:2008fd,Tsang:2012se,Lattimer:2014scr}.
According to the recent Bayesian approach with correlated uncertainties of the infinite-matter EoS derived from chiral effective field theory, $E_{\rm sym}$ and  $L$ are predicted to be $E_{\rm sym}=31.7\pm1.1$ MeV and $L=59.8\pm4.1$ MeV at the nuclear saturation density \citep{Drischler:2020hwi}.
Concurrently, it is possible to give constraints on those physical quantities using some astrophysical information on neutron stars, such as the radius measurements from {\it NICER} and {\it XMM-Newton} data \citep{Miller:2021qha}, and the tidal deformability due to gravitational wave (GW) signals from the binary neutron star merger, GW170817 \citep{LIGOScientific:2018cki,LIGOScientific:2018hze}.

From the viewpoint of theoretical studies on $E_{\rm sym}$ and $L$, the isovector, Lorentz-scalar ($\bm{\delta}$) meson can be included in the RMF calculations if we remind that the OBE potential was successful for understanding nuclear interactions \citep{Kubis:1997ew,Hofmann:2000vz,Liu:2001iz}.
The $\delta$ meson, however, has been claimed to be less important than the $\rho$ meson so far to reproduce the properties of asymmetric nuclear matter because of its small impact on the nuclear EoS even at high densities \citep{Greco:2002dv,Bunta:2004ej,Menezes:2004vr}.
On the other hand, it has been realized that the $\delta$ meson strongly affects the proton fraction in neutron-star matter and hence the cooling process of a neutron star, using the density-dependent RMF model which includes the $\sigma$, $\omega^{\mu}$, $\bm{\delta}$, and $\bm{\rho}^{\mu}$ mesons with density-dependent meson-nucleon couplings \citep{Roca-Maza:2011alv,Wang:2014jmr,Typel:2020ozc}.
Furthermore, a new type of scalar-meson mixing, e.g., $\sigma\bm{\delta}^{2}$ and $\sigma^{2}\bm{\delta}^{2}$, has been recently introduced in the RMF model, and it gives a large influence not only on $E_{\rm sym}$ but also on $L$ \citep{Zabari:2018tjk,Zabari:2019clk,Kubis:2020ysv}.

In the present study, using the RMF model, we investigate the $\delta$-meson effect on the properties of isospin-asymmetric nuclear matter.
Then, our results are compared with the experimental constraints on $E_{\rm sym}$ and $L$ as well as the recent data from astrophysical observations.
In particular, we study the influence of isoscalar- and isovector-meson mixing, $\sigma^{2}\bm{\delta}^{2}$ and $\omega_{\mu}\omega^{\mu}\bm{\rho}_{\nu}\bm{\rho}^{\nu}$, on the density dependence of $E_{\rm sym}$ and the EoS for neutron stars.

This paper is organized as follows.
A brief review of the RMF model with several species of nonlinear couplings is provided in Section \ref{sec:framework}.
Numerical results and detailed discussions concerning features of isospin-asymmetric nuclear and neutron-star matter are presented in Section \ref{sec:results}.
Finally, we give a summary in Section \ref{sec:summary}.

\section{Theoretical Framework} \label{sec:framework}

We employ the RMF model based on quantum hadrodynamics \citep{Walecka:1974qa,Serot:1984ey}.
The Lagrangian density, $\mathcal{L}$, includes the fields of nucleons ($N=p,n$) and mesons.
We here introduce four mesons: $\sigma$, $\omega^{\mu}$, $\bm{\delta}$, and $\bm{\rho}^{\mu}$.
The Lagrangian density is thus chosen to be
\begin{align}
  \mathcal{L}
  & = \sum_{N=p,n}\bar{\psi}_{N}\bigl[i\gamma_{\mu}\partial^{\mu}
    - \left(M_{N}-g_{\sigma}\sigma-g_{\delta}\bm{\delta}\cdot\bm{\tau}_{N}\right)
    \nonumber \\
  & - g_{\omega}\gamma_{\mu}\omega^{\mu}
    - g_{\rho}\gamma_{\mu}\bm{\rho}^{\mu}\cdot\bm{\tau}_{N}\bigr]\psi_{N}
    - U_{\rm NL}(\sigma,\omega^{\mu},\bm{\delta},\bm{\rho}^{\mu})
    \nonumber \\
  & + \frac{1}{2}\left(\partial_{\mu}\sigma\partial^{\mu}\sigma-m_{\sigma}^{2}\sigma^{2}\right)
    + \frac{1}{2}m_{\omega}^{2}\omega_{\mu}\omega^{\mu}-\frac{1}{4}W_{\mu\nu}W^{\mu\nu}
    \nonumber \\
  & + \frac{1}{2}\left(\partial_{\mu}\bm{\delta}\cdot\partial^{\mu}\bm{\delta}-m_{\delta}^{2}\bm{\delta}^{2}\right)
    + \frac{1}{2}m_{\rho}^{2}\bm{\rho}_{\mu}\cdot\bm{\rho}^{\mu}-\frac{1}{4}\bm{R}_{\mu\nu}\cdot\bm{R}^{\mu\nu},
    \label{eq:Lagrangian}
\end{align}
where $\psi_{N}$ is the nucleon field, $\bm{\tau}_{N}$ is its isospin matrix, $W_{\mu\nu}=\partial_{\mu}\omega_{\nu}-\partial_{\nu}\omega_{\mu}$, and $\bm{R}_{\mu\nu}=\partial_{\mu}\bm{\rho}_{\nu}-\partial_{\nu}\bm{\rho}_{\mu}$.
The meson-nucleon coupling constants are respectively denoted by $g_{\sigma}$, $g_{\omega}$, $g_{\delta}$, and $g_{\rho}$.
Additionally, a nonlinear potential in Equation \eqref{eq:Lagrangian} is supplemented as follows:
\begin{align}
  U_{NL}(\sigma,\omega^{\mu},\bm{\delta},\bm{\rho}^{\mu})
  & = \frac{1}{3}g_{2}\sigma^{3} + \frac{1}{4}g_{3}\sigma^{4}
    - \Lambda_{s}\left(g_{\sigma}^{2}\sigma^{2}\right)\left(g_{\delta}^{2}\bm{\delta}^{2}\right)
    \nonumber \\
  & - \Lambda_{v}\left(g_{\omega}^{2}\omega_{\mu}\omega^{\mu}\right)\left(g_{\rho}^{2}\bm{\rho}_{\mu}\cdot\bm{\rho}^{\mu}\right).
    \label{eq:NL-potential}
\end{align}
The first and second terms in Equation \eqref{eq:NL-potential} are introduced to obtain a  quantitative description of ground-state properties for symmetric nuclear matter \citep{Boguta:1977xi}.
In contrast, the third and forth terms in Equation \eqref{eq:NL-potential} only affect the characteristics of isospin-asymmetric nuclear matter \citep{ToddRutel:2005yzo,Miyatsu:2013yta}.
Although it is possible to consider the linear type of $\sigma$-$\delta$ interaction, $\sigma\bm{\delta}^{2}$, based on the Lorentz covariance of $\mathcal{L}$, only the quadratic interaction, $\sigma^{2}\bm{\delta}^{2}$, is here considered because it provides a large impact on the density dependence of $E_{\rm sym}$ \citep{Zabari:2018tjk,Zabari:2019clk}.
Therefore, the potential involves four coupling constants and mixing parameters, $g_{2}$, $g_{3}$, $\Lambda_{s}$, and $\Lambda_{v}$.
For convenience, we hereafter use $\Lambda_{\sigma\delta}$ and $\Lambda_{\omega\rho}$ instead of $\Lambda_{s}$ and $\Lambda_{v}$, i.e., $\Lambda_{\sigma\delta}\equiv\Lambda_{s}g_{\sigma}^{2}g_{\delta}^{2}$ and $\Lambda_{\omega\rho}\equiv\Lambda_{v}g_{\omega}^{2}g_{\rho}^{2}$.
In the present study, the nucleon and meson masses in vacuum are taken as follows: $M_{N}=939$ MeV, $m_{\sigma}=500$ MeV, $m_{\omega}=783$ MeV, $m_{\delta}=983$ MeV, and $m_{\rho}=770$ MeV.

In RMF approximation, the meson fields are replaced by the mean-field values: $\bar{\sigma}$, $\bar{\omega}$, $\bar{\delta}$, and $\bar{\rho}$.
Then, the effective nucleon mass, $M_{N}^{\ast}$, is simply expressed as
\begin{equation}
  M_{p \choose n}^{\ast}(\bar{\sigma},\bar{\delta}) = M_{N} - g_{\sigma}\bar{\sigma} \mp g_{\delta}\bar{\delta}.
  \label{eq:effectivemass}
\end{equation}
The equations of motion for the meson fields in uniform matter are thus given by
%
%
\begin{align}
  \left(m_{\sigma}^{2}+g_{2}\bar{\sigma}+g_{3}\bar{\sigma}^{2}-2\Lambda_{\sigma\delta}\bar{\delta}^{2}\right)\bar{\sigma}
  & = g_{\sigma}\left(\rho_{p}^{s}+\rho_{n}^{s}\right),
    \label{eq:EoM-sigma} \\
  \left(m_{\omega}^{2}+2\Lambda_{\omega\rho}\bar{\rho}^{2}\right)\bar{\omega}
  & = g_{\omega}\left(\rho_{p}+\rho_{n}\right),
    \label{eq:EoM-omega} \\
  \left(m_{\delta}^{2}-2\Lambda_{\sigma\delta}\bar{\sigma}^{2}\right)\bar{\delta}
  & = g_{\delta}\left(\rho_{p}^{s}-\rho_{n}^{s}\right),
    \label{eq:EoM-delta} \\
  \left(m_{\rho}^{2}+2\Lambda_{\omega\rho}\bar{\omega}^{2}\right)\bar{\rho}
  & = g_{\rho}\left(\rho_{p}-\rho_{n}\right),
    \label{eq:EoM-rho}
\end{align}
where the scalar density, $\rho_{N}^{s}$, and the baryon density, $\rho_{N}$, read
\begin{align}
  \rho_{N}^{s}
  & = \frac{1}{\pi^{2}}\int_{0}^{k_{F_{N}}}dk~k^{2}\frac{M_{N}^{\ast}(\bar{\sigma},\bar{\delta})}{\sqrt{k^{2}+M_{N}^{\ast2}(\bar{\sigma},\bar{\delta})}},
    \label{eq:scalar-density} \\
  \rho_{N}
  & = \frac{k^{3}_{F_{N}}}{3\pi^{2}},
    \label{eq:baryon-density}
\end{align}
with $k_{F_N}$ being the Fermi momentum for $N$.

With the self-consistent calculations of the meson fields given in Equations \eqref{eq:EoM-sigma}--\eqref{eq:EoM-rho}, the energy density, $\varepsilon$, and pressure, $P$, in nuclear matter are given by
\begin{align}
  \varepsilon
  & = \sum_{N}\frac{1}{\pi^{2}}\int_{0}^{k_{F_{N}}}dk~k^{2}\sqrt{k^{2}+M_{N}^{\ast2}(\bar{\sigma},\bar{\delta})}
    \nonumber \\
  & + \frac{1}{2}\left(m_{\sigma}^{2}\bar{\sigma}^{2}+m_{\omega}^{2}\bar{\omega}^{2}+m_{\delta}^{2}\bar{\delta}^{2}+m_{\rho}^{2}\bar{\rho}^{2}\right)
    \nonumber \\
  & + \frac{1}{3}g_{2}\bar{\sigma}^{3}
    + \frac{1}{4}g_{3}\bar{\sigma}^{4}
    - \Lambda_{\sigma\delta}\bar{\sigma}^{2}\bar{\delta}^{2}
    + 3\Lambda_{\omega\rho}\bar{\omega}^{2}\bar{\rho}^{2},
    \label{eq:engy-density} \\
  P
  & = \frac{1}{3}\sum_{N}\frac{1}{\pi^{2}}\int_{0}^{k_{F_{N}}}dk~\frac{k^{4}}{\sqrt{k^{2}+M_{N}^{\ast2}(\bar{\sigma},\bar{\delta})}}
    \nonumber \\
  & - \frac{1}{2}\left(m_{\sigma}^{2}\bar{\sigma}^{2}-m_{\omega}^{2}\bar{\omega}^{2}+m_{\delta}^{2}\bar{\delta}^{2}-m_{\rho}^{2}\bar{\rho}^{2}\right)
    \nonumber \\
  & - \frac{1}{3}g_{2}\bar{\sigma}^{3}
    - \frac{1}{4}g_{3}\bar{\sigma}^{4}
    + \Lambda_{\sigma\delta}\bar{\sigma}^{2}\bar{\delta}^{2}
    + \Lambda_{\omega\rho} \bar{\omega}^{2} \bar{\rho}^{2}.
    \label{eq:pressure}
\end{align}

According to the Hugenholtz–-Van Hove (HVH) theorem \citep{Czerski:2002pz,Cai:2012en}, $E_{\rm sym}$ is generally divided into the kinetic and potential terms as $E_{\rm sym}=E_{\rm sym}^{\rm kin}+E_{\rm sym}^{\rm pot}$ \citep{Miyatsu:2020vzi}, and they are respectively given by
\begin{align}
  E_{\rm sym}^{\rm kin}
  & = \frac{1}{6}\frac{k_{F}^{2}}{\sqrt{k_{F}^{2}+M_{F}^{\ast2}}},
    \label{eq:Esym-kin} \\
  E_{\rm sym}^{\rm pot}
  & = E_{\rm sym}^{\rho}+E_{\rm sym}^{\delta}
    \nonumber \\
  & = \frac{1}{2}\frac{g_{\rho}^{2}}{m_{\rho}^{\ast2}}\rho_{B}
    - \frac{1}{2}\frac{g_{\delta}^{2}}{m_{\delta}^{\ast2}}\rho_{B}\left(\frac{M_{F}^{\ast2}}{k_{F}^{2}+M_{F}^{\ast2}}\right)I_{F},
    \label{eq:Esym-pot}
\end{align}
at $\rho_p = \rho_n$, namely $k_{F}=k_{F_{p}}=k_{F_{n}}$ and $M_{F}^{\ast}=M_{p}^{\ast}=M_{n}^{\ast}$.
Here, the effective meson masses in matter are defined as $m_{\delta}^{\ast2}=m_{\delta}^{2}-2\Lambda_{\sigma\delta}\bar{\sigma}^{2}$ and $m_{\rho}^{\ast2}=m_{\rho}^{2}+2\Lambda_{\omega\rho}\bar{\omega}^{2}$, and
\begin{equation}
  I_{F}= \left[1+3\frac{g_{\delta}^{2}}{m_{\delta}^{\ast2}}\left(\frac{\rho_{B}^{s}}{M_{F}^{\ast}}-\frac{\rho_{B}}{\sqrt{k_{F}^{2}+M_{F}^{\ast2}}}\right)\right]^{-1},
\end{equation}
where the total scalar and baryon densities are written as $\rho_{B}^{s}=\rho_{p}^{s}+\rho_{n}^{s}$ and $\rho_{B}=\rho_{p}+\rho_{n}$, respectively.

\section{Numerical Results} \label{sec:results}

In total, there are eight coupling constants which have to be determined in Equation \eqref{eq:Lagrangian}.
They can be classified into two categories.
One is the coupling constants which affect the saturation properties of symmetric nuclear matter.
The other is the couplings related to the nature of isospin-asymmetric nuclear matter.
In order to determine the coupling constants concerning symmetric nuclear matter at the saturation density, $\rho_{0}$, we use the recent constraints from terrestrial experiments and astrophysical observations of neutron stars as follows \citep{Choi:2020eun}: the binding energy per nucleon ($E_{0}=-16.0$ MeV), the nuclear incompressibility ($K_{0}=230$ MeV), and the effective nucleon mass ($M_{N}^{\ast}/M_{N}=0.65$) at $\rho_{0}=0.16$ fm$^{-3}$.
Finally, we get the values, $g_{\sigma}=9.22$, $g_{\omega}=11.35$, $g_{2}=13.08$ fm$^{-1}$, and $g_{3}=-31.60$.

\begin{table}[t]
  \caption{\label{tab:CCs}%
    Coupling constants for isovector mesons and mixing parameters.
    The value of $g_{\delta}^{2}$, which is taken from the OBE potential \citep{Machleidt:1989tm}, is denoted by $A$, $B$, or $C$.
    For details, see the text.}
  \begin{ruledtabular}
    \begin{tabular}{ccrr}
      $g_{\delta}^{2}/4\pi$        & $g_{\rho}^{2}/4\pi$ & $\Lambda_{\sigma\delta}$ & $\Lambda_{\omega\rho}$ \\
      \colrule
      0                            &                2.41 &                       -- &               $654.13$ \\
      \colrule
      \multirow{5}{*}{1.3 ($A$)}   &                3.08 &                   $-100$ &               $493.71$ \\
      \                            &                3.01 &                    $-50$ &               $431.48$ \\
      \                            &                2.91 &                      $0$ &               $355.53$ \\
      \                            &                2.77 &                     $50$ &               $263.92$ \\
      \                            &                2.59 &                    $100$ &               $155.49$ \\
      \colrule
      \multirow{5}{*}{2.488 ($B$)} &                3.76 &                   $-100$ &               $433.29$ \\
      \                            &                3.67 &                    $-50$ &               $357.70$ \\
      \                            &                3.54 &                      $0$ &               $271.21$ \\
      \                            &                3.39 &                     $50$ &               $173.77$ \\
      \                            &                3.19 &                    $100$ &                $66.18$ \\
      \colrule
      \multirow{5}{*}{4.722 ($C$)} &                5.07 &                   $-100$ &               $382.25$ \\
      \                            &                4.97 &                    $-50$ &               $298.92$ \\
      \                            &                4.83 &                      $0$ &               $208.39$ \\
      \                            &                4.66 &                     $50$ &               $111.38$ \\
      \                            &                4.46 &                    $100$ &                 $9.04$ \\
      \colrule
      \multirow{5}{*}{10}          &                8.27 &                   $-100$ &               $347.85$ \\
      \                            &                8.15 &                    $-50$ &               $260.48$ \\
      \                            &                8.00 &                      $0$ &               $169.28$ \\
      \                            &                7.83 &                     $50$ &                $74.94$ \\
      \                            &                7.62 &                    $100$ &               $-21.71$ \\
    \end{tabular}
  \end{ruledtabular}
\end{table}
On the other hand, the coupling constants for isovector mesons and the mixing parameters, $g_{\delta}$, $g_{\rho}$, $\Lambda_{\sigma\delta}$, and $\Lambda_{\omega\rho}$, are fixed by the properties of asymmetric nuclear matter.
We here set $E_{\rm sym}=32.0$ MeV and $L=50$ MeV at $\rho_{0}$ to explain the recent astrophysical observations \citep{Choi:2020eun}.
Moreover, if the acceptable coupling of $g_{\delta}$ based on the OBE potential \citep{Machleidt:1989tm} and the recent result of $\Lambda_{\sigma\delta}$ given in \citet{Zabari:2018tjk,Zabari:2019clk} are taken into consideration, they are supposed to be varied in the ranges of $0\le g_{\delta}^{2}/4\pi\le10$ and $-100\le\Lambda_{\sigma\delta}\le100$, respectively.
Once $g_{\delta}$ and $\Lambda_{\sigma\delta}$ are fixed in those ranges, it is possible to determine the others ($g_{\rho}$ and $\Lambda_{\omega\rho}$) by adopting the given $E_{\rm sym}$ and $L$.
The coupling constants related to isospin-asymmetric features in the present study are listed in Table \ref{tab:CCs}.

\begin{figure}
  \plotone{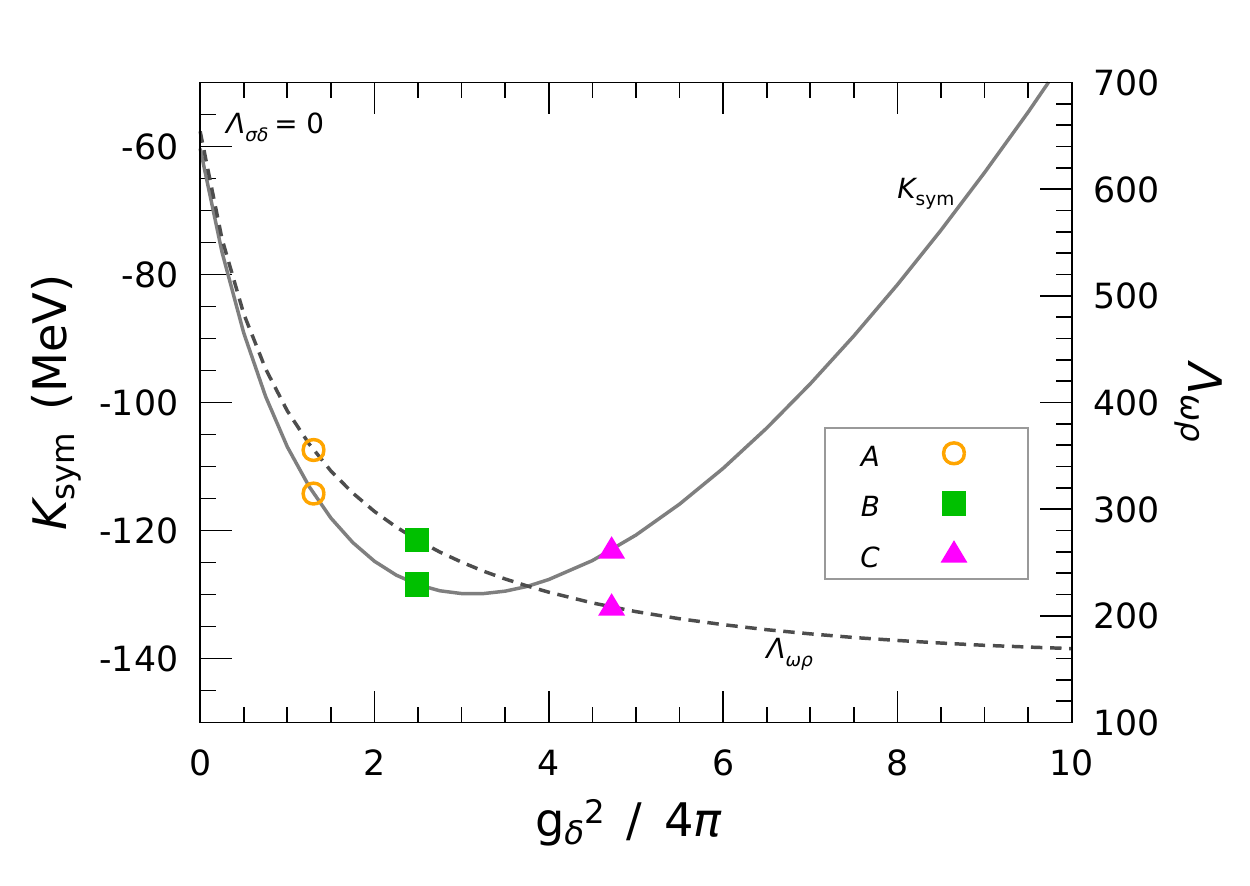}
  \caption{\label{fig:Ksym-gd2}
    Correlations between $K_{\rm sym}$ and the coupling constants for $\Lambda_{\sigma\delta}=0$.
    The solid (dashed) line denotes $K_{\rm sym}$ ($\Lambda_{\omega\rho}$).
    We also mark the points which correspond to the results obtained from potential $A$, $B$, and $C$ \citep{Machleidt:1989tm}.}
\end{figure}
\begin{figure}
  \plotone{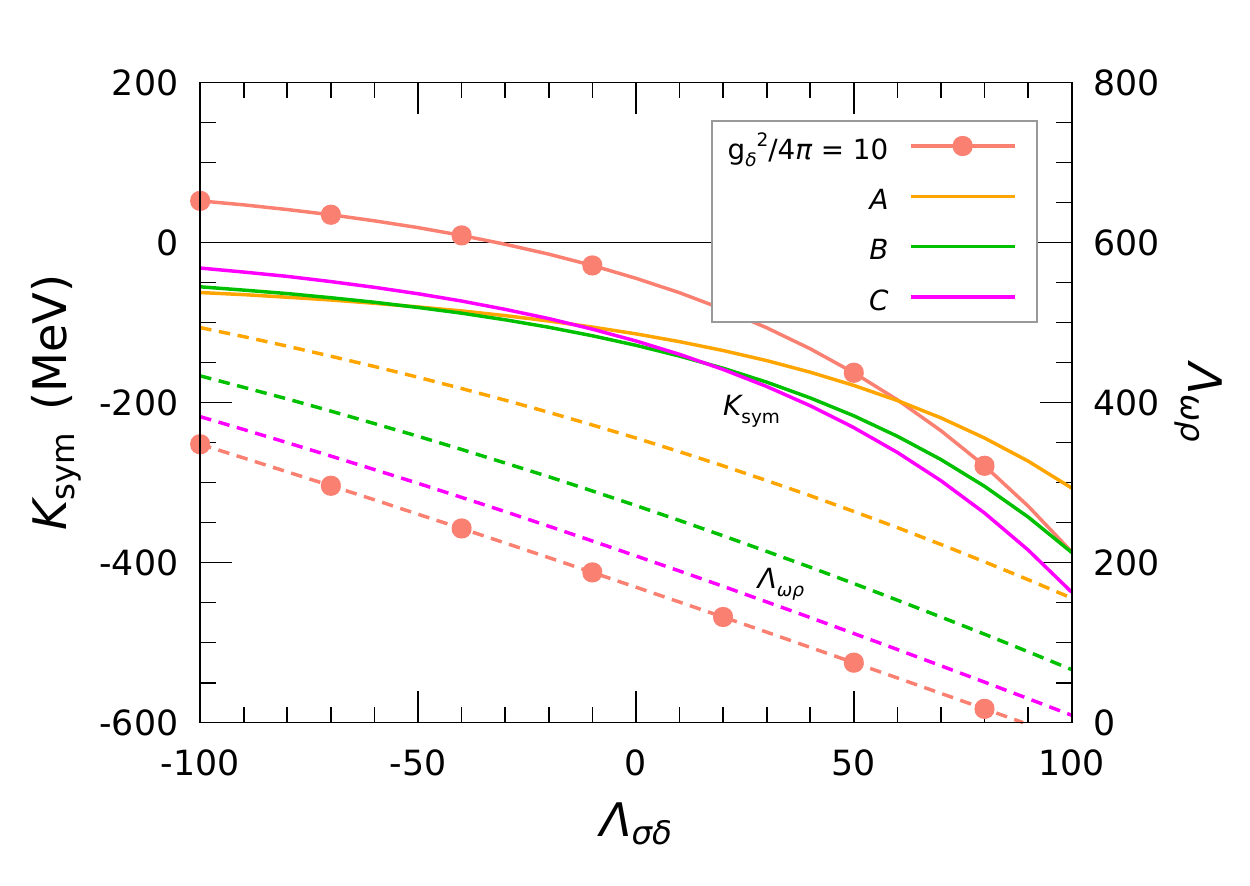}
  \caption{\label{fig:Ksym-Lsd}
    Same as Figure \ref{fig:Ksym-gd2} but with the $\sigma$-$\delta$ mixing.
    Each line is calculated by using the fixed $g_{\delta}^{2}$ given in Table \ref{tab:CCs}.}
\end{figure}
We present the correlations between the curvature parameter of nuclear symmetry energy, $K_{\rm sym}$, and the related coupling constants in Figures \ref{fig:Ksym-gd2} and \ref{fig:Ksym-Lsd}.
In order to focus on the $\delta$-meson effect, the result without the $\sigma$-$\delta$ mixing is shown in Figure \ref{fig:Ksym-gd2}.
It is found that $K_{\rm sym}$ has the minimum point around the result of potential $B$, while, as $g_{\delta}^{2}$ increases, $\Lambda_{\omega\rho}$ first decreases rapidly and then becomes almost constant.
In contrast, we show the result including the $\sigma$-$\delta$ mixing in Figure \ref{fig:Ksym-Lsd}.
We have found that $K_{\rm sym}$ varies in the range of $-450\leq K_{\rm sym}\mathrm{(MeV)}\leq50$, and that $\Lambda_{\omega\rho}$ becomes small as $\Lambda_{\sigma\delta}$ increases in all the cases.
It implies that, as explained in \citet{Zabari:2018tjk}, the $\sigma$-$\delta$ mixing can partly take on a role of the $\omega$-$\rho$ mixing in describing the properties of asymmetric nuclear matter, such as $E_{\rm sym}$, $L$, and $K_{\rm sym}$.

\begin{figure}
  \plotone{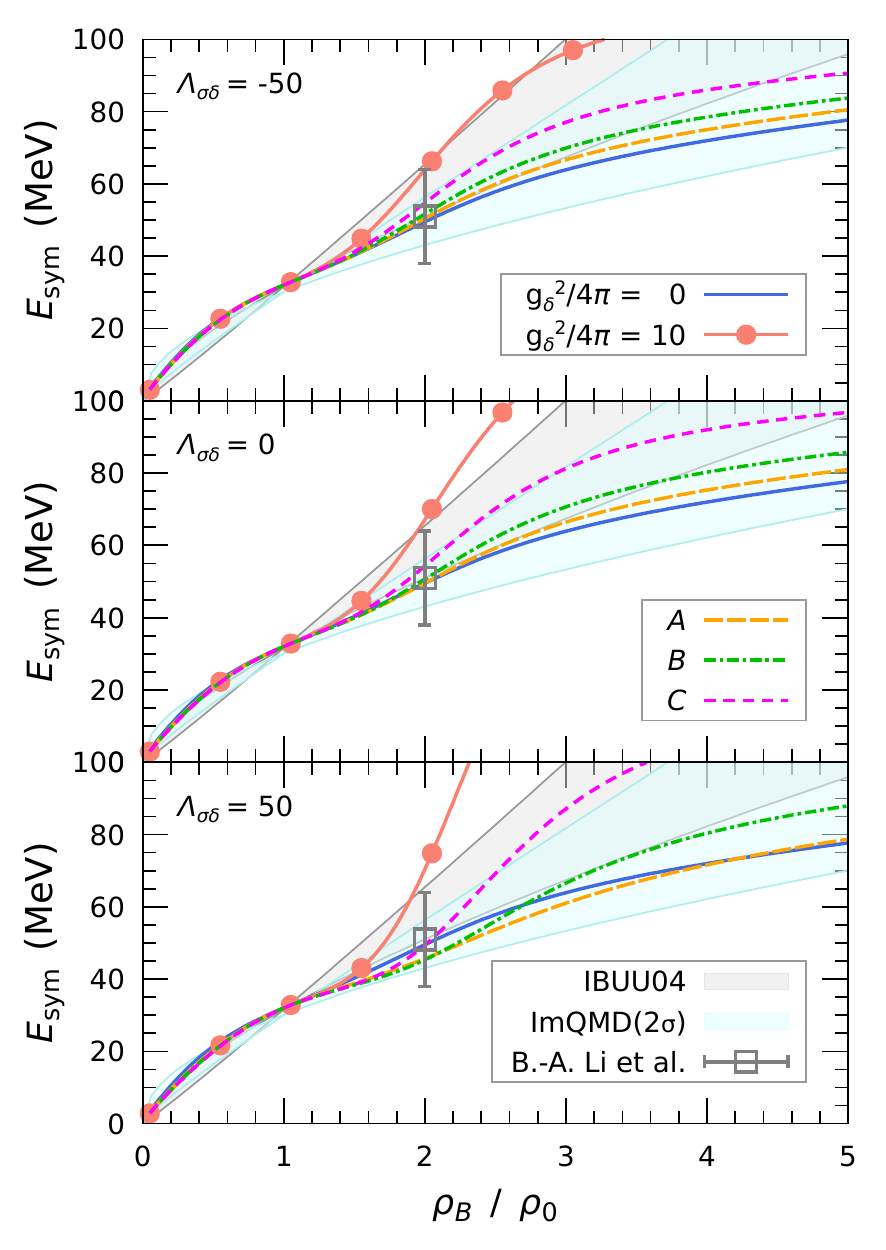}
  \caption{\label{fig:Esym}
    Nuclear symmetry energy, $E_{\rm sym}$, as a function of $\rho_{B}/\rho_{0}$.
    The results are calculated with the fixed $g_{\delta}^2$ given in Table \ref{tab:CCs}.
    The top (middle) [bottom] panel is for the case of $\Lambda_{\sigma\delta}=-50$ ($0$) [$50$].
    For details, see the text.}
\end{figure}
\begin{figure}
  \plotone{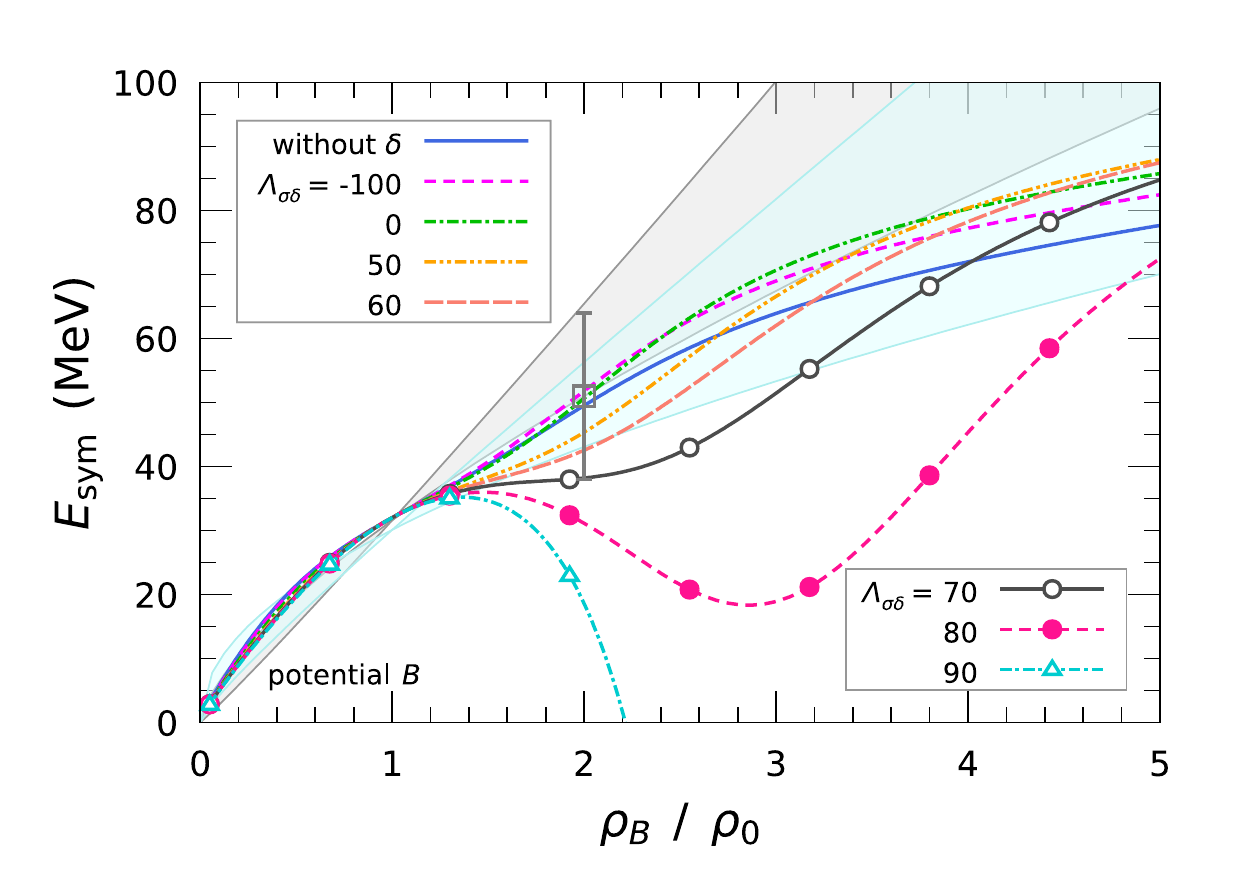}
  \caption{\label{fig:Esym-B}
    Same as Figure \ref{fig:Esym} but with the $\sigma$-$\delta$ mixing.
    The coupling constants in potential $B$ are used.}
\end{figure}
The density dependence of $E_{\rm sym}$ is depicted in Figure \ref{fig:Esym}.
The constraints from analyses of heavy-ion collision data using the isospin-dependent Boltzmann-Uehling-Uhlenbec (IBUU04) and improved quantum molecular dynamics (ImQMD) transport models are presented \citep{Tsang:2008fd,Chen:2004si,Li:2005jy}.
We also show the recent constraint on the magnitude of $E_{\rm sym}$ at $2\rho_{0}$, i.e., $E_{\rm sym}(2\rho_{0})\simeq51\pm13$ MeV at a $68\%$ confidence level, from nine new analyses of neutron-star observables since GW170817 \citep{Li:2021thg}.
It shows that $E_{\rm sym}$ is sensitive to $g_{\delta}^{2}$ above $\rho_{0}$, i.e., as $g_{\delta}^{2}$ increases, it becomes large at high densities in all the cases.
Thus, the $\delta$ meson enhances $E_{\rm sym}$ in dense nuclear matter.
However, in the case of $g_{\delta}^{2}/4\pi=10$, it is too large to explain the $E_{\rm sym}(2\rho_{0})$ restriction.
On the other hand, for $-50 \leq \Lambda_{\sigma\delta} \leq 50$, $E_{\rm sym}$ in potential $A$, $B$ or $C$, lies in the region of the constraints from the IBUU04 and/or ImQMD transport model.

For the sake of studying the $\sigma$-$\delta$ mixing effect in detail, $E_{\rm sym}$ with the coupling constants in potential $B$ is given in Figure \ref{fig:Esym-B}.
The $\sigma$-$\delta$ mixing has a weak influence on $E_{\rm sym}$ below $\rho_{0}$.
However, as explained in \citet{Zabari:2018tjk}, the $\sigma$-$\delta$ mixing strongly affects $E_{\rm sym}$ above $\rho_{0}$.
For example, the $\sigma$-$\delta$ mixing reduces $E_{\rm sym}$ at high densities, and then partly cancels the enhancement due to the $\delta$-$N$ interaction.
Furthermore, for $70<\Lambda_{\sigma\delta}<80$, $E_{\rm sym}$ has an inflection point above $\rho_{0}$, and the dip then appears around $2\rho_{0}$--$3\rho_{0}$.
For $\Lambda_{\sigma\delta}\geq90$, $E_{\rm sym}$ becomes negative at high densities.
To satisfy the experimental constraints from heavy-ion collision data and the analytical result obtained from neutron-star observations, we find that $\Lambda_{\sigma\delta}$ should be less than $60$.

\begin{figure}
  \plotone{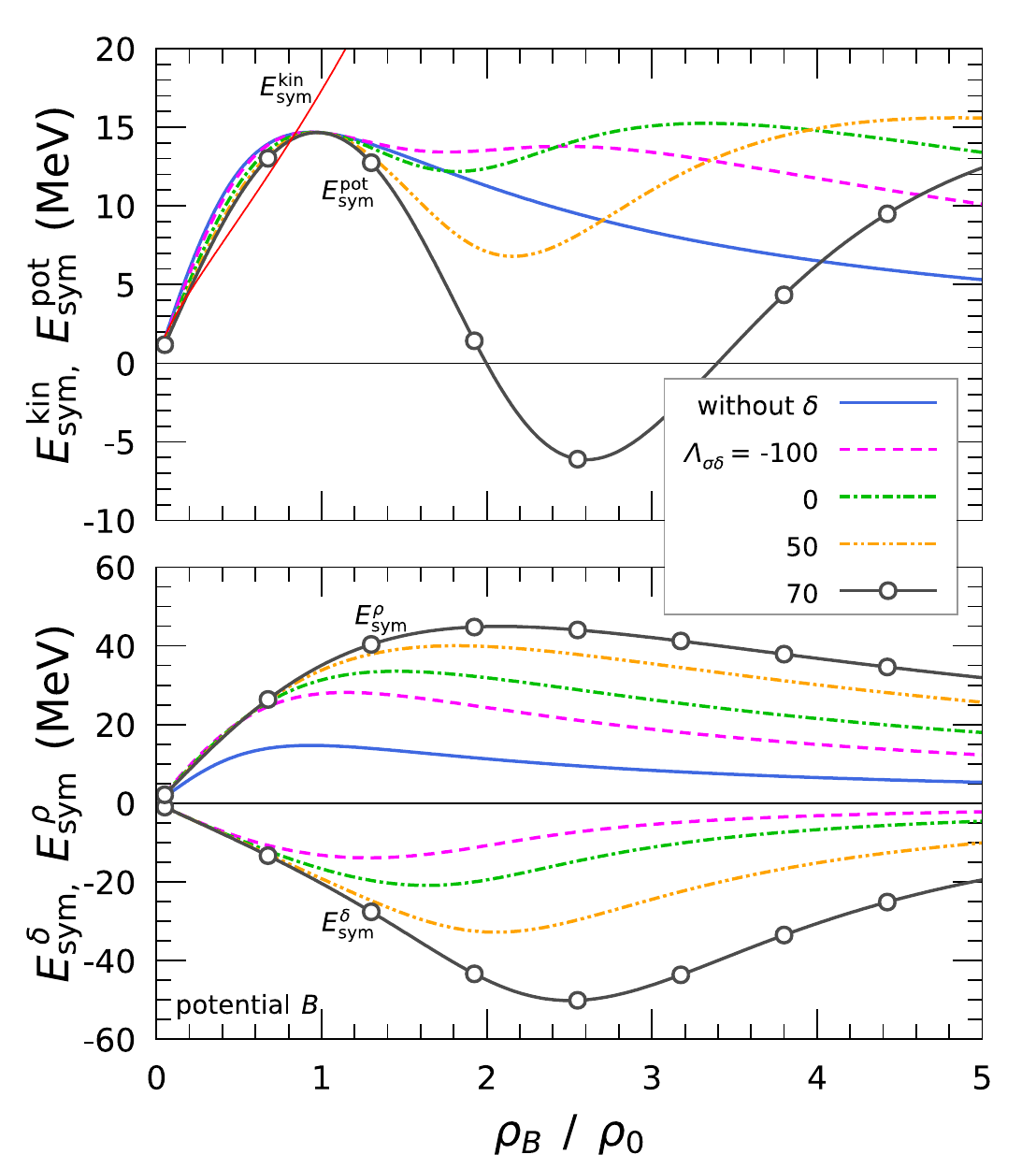}
  \caption{\label{fig:Esym-detail}
    Contents of $E_{\rm sym}$ as a function of $\rho_{B}/\rho_{0}$.
    The coupling constants in potential $B$ are used.
    The kinetic and potential terms, $E_{\rm sym}^{\rm kin}$ and $E_{\rm sym}^{\rm pot}$, are shown in the upper panel.
    The meson contributions, $E_{\rm sym}^{\delta}$ and $E_{\rm sym}^{\rho}$, are presented in the lower panel.}
\end{figure}
In the upper panel of Figure \ref{fig:Esym-detail}, the kinetic and potential terms of $E_{\rm sym}$ given in Equations \eqref{eq:Esym-kin} and \eqref{eq:Esym-pot} are presented.
Because $E_{\rm sym}^{\rm kin}$ is calculated by the Fermi momentum and the effective nucleon mass at $\rho_p=\rho_n$ (see Equation \eqref{eq:Esym-kin}), it is common to all the cases.
On the contrary, $E_{\rm sym}^{\rm pot}$ shows the unique behavior above $\rho_{0}$.
When only the $\rho$ meson and its quadratic interaction, $\omega_{\mu}\omega^{\mu}\bm{\rho}_{\nu}\bm{\rho}^{nu}$, are considered, $E_{\rm sym}^{\rm pot}$ shows the maximum point around $\rho_{0}$ and decreases monotonically as the density increases.
In contrast, $E_{\rm sym}^{\rm pot}$ becomes stiff at high densities when the $\delta$ meson is considered.
For $\Lambda_{\sigma\delta}=0$, $E_{\rm sym}^{\rm pot}$ reaches practically plateau above $\rho_{0}$.
Moreover, for $\Lambda_{\sigma\delta}\geq50$, the rapid reduction occurs above $\rho_{0}$, and then $E_{\rm sym}^{\rm pot}$ turns to grow as the density increases.
Thus, $E_{\rm sym}$ becomes temporarily soft around $2\rho_{0}$--$3\rho_{0}$, as already seen in Figure \ref{fig:Esym-B}.

The meson contributions to $E_{\rm sym}^{\rm pot}$, which are composed of $E_{\rm sym}^{\delta}$ and $E_{\rm sym}^{\rho}$ defined in Equation \eqref{eq:Esym-pot}, are given in the lower panel of Figure \ref{fig:Esym-detail}.
We note that $E_{\rm sym}^{\delta}$ ($E_{\rm sym}^{\rho}$) contributes to $E_{\rm sym}^{\rm pot}$ negatively (positively).
When the absolute value of $E_{\rm sym}^{\delta}$ is larger than that of $E_{\rm sym}^{\rho}$, $E_{\rm sym}^{\rm pot}$ has a rapid reduction, and, accordingly, $E_{\rm sym}$ has the dip, as already shown in Figure \ref{fig:Esym-B} and in the upper panel of Figure \ref{fig:Esym-detail}.

\begin{figure}
  \plotone{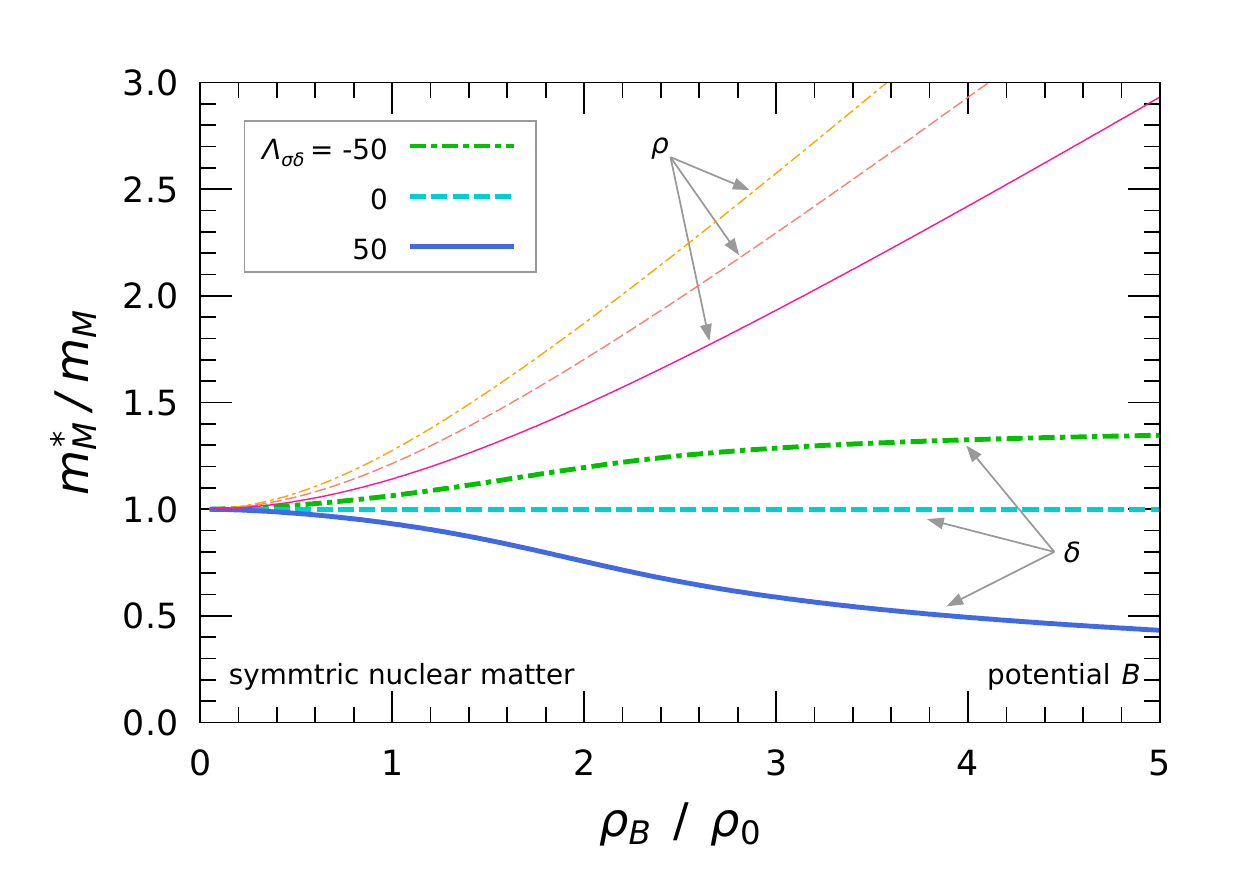}
  \caption{\label{fig:Emass-meson-B}
    Effective mass ratio of isovector mesons in symmetric nuclear matter as a function of $\rho_{B}/\rho_{0}$.
    The coupling constants in potential $B$ are used.
    The thick (thin) lines correspond to the mass of the $\delta$ ($\rho$) meson.
    The solid (dashed) [dot--dashed] lines are for $\Lambda_{\sigma\delta}=50$ ($0$) [$-50$].}
\end{figure}
In Figure \ref{fig:Emass-meson-B}, we present the effective masses of isovector mesons, $m_{\delta}^{\ast}$ and $m_{\rho}^{\ast}$, in symmetric nuclear matter.
Because $m_{\delta}^{\ast}$ and $m_{\rho}^{\ast}$ respectively couple with the isoscalar-meson fields, $\bar{\sigma}$ and $\bar{\omega}$, through the mixing, they change significantly at high densities.
In particular, $m_{\rho}^{\ast}$ varies remarkably due to the large $\Lambda_{\omega\rho}$, compared with the case of $m_{\delta}^{\ast}$.
Because $m_{\delta}^{\ast}$ and $m_{\rho}^{\ast}$ become small as $\Lambda_{\sigma\delta}$ increases, $\left|E_{\rm sym}^{\delta}\right|$ and $\left|E_{\rm sym}^{\rho}\right|$ become large with increasing $\Lambda_{\sigma\delta}$ (see the lower panel of Figure \ref{fig:Esym-detail} and Equation \eqref{eq:Esym-pot}).

\begin{figure}
  \plotone{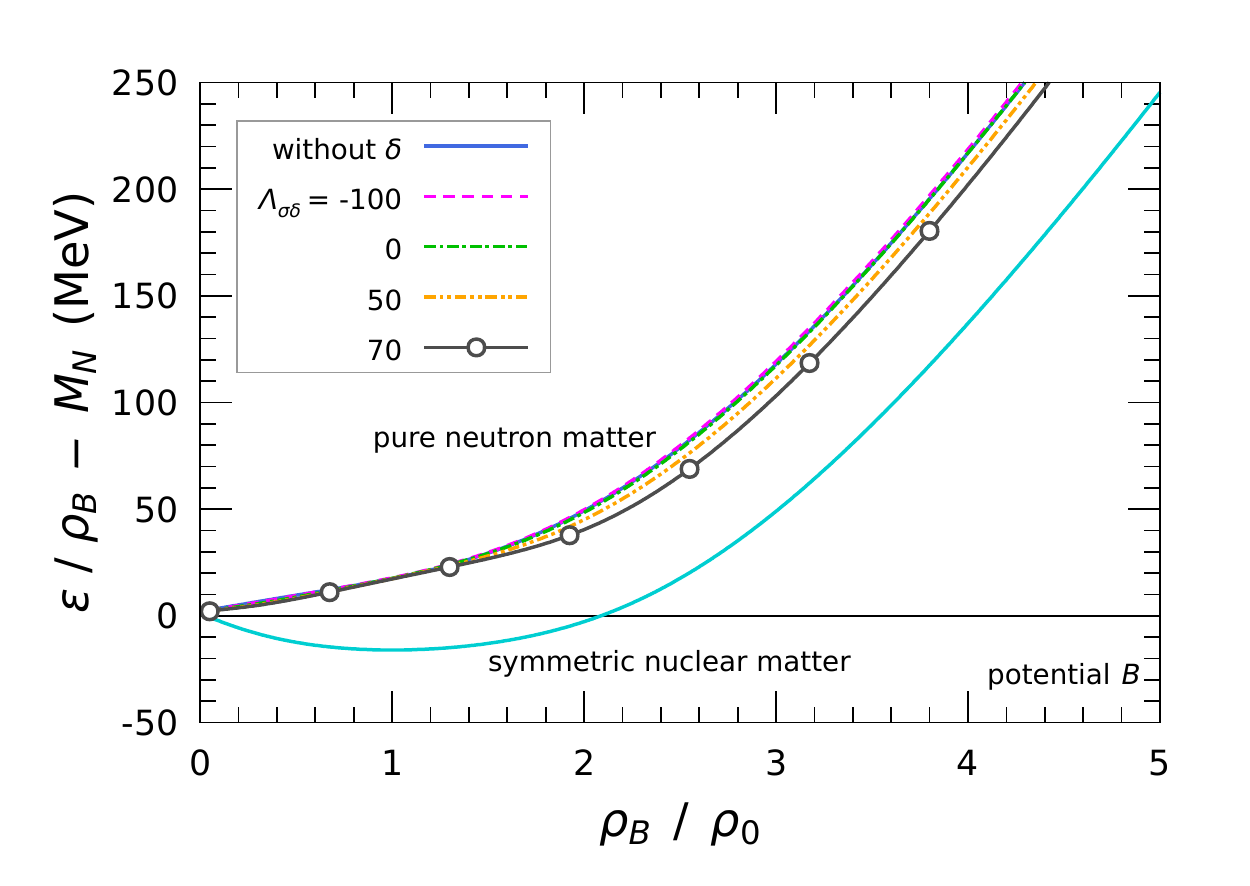}
  \caption{\label{fig:Bind-engy}
    Binding energy per nucleon, $E_{B}$, in symmetric nuclear and pure neutron matter as a function of $\rho_{B}/\rho_{0}$.
    The coupling constants in potential $B$ are used.}
\end{figure}
The binding energy per nucleon, $E_{B} = \varepsilon /\rho_B - M_N$, is illustrated in Figure \ref{fig:Bind-engy}.
As in the case of $E_{\rm sym}^{\rm kin}$, the $\delta$ meson has no influence on $E_{B}$ in symmetric nuclear matter.
In addition, the $\sigma$-$\delta$ mixing for $\Lambda_{\sigma\delta} < 0$ rarely affects $E_{B}$ even in pure neutron matter.
In contrast, for $\Lambda_{\sigma\delta}>0$, $E_{B}$ is suppressed by the $\sigma$-$\delta$ mixing at high densities.
It is thus found that the positive $\sigma$-$\delta$ mixing decreases the energy difference between pure neutron and symmetric nuclear matter as the density increases, and that it consequently gives the softer $E_{\rm sym}$ at high densities as shown in Figure \ref{fig:Esym-B}.
In other words, if one considers the $\sigma$-$\delta$ mixing,  in the present calculation, there is still room for employing the higher $E_{\rm sym}$ and $L$ at $\rho_0$, which are recently suggested by PREX-II data \citep{PREX:2021umo,Piekarewicz:2021jte,Reed:2021nqk}, in determining the coupling constants.
We will again discuss this at the end of this section.

\begin{figure}
  \plotone{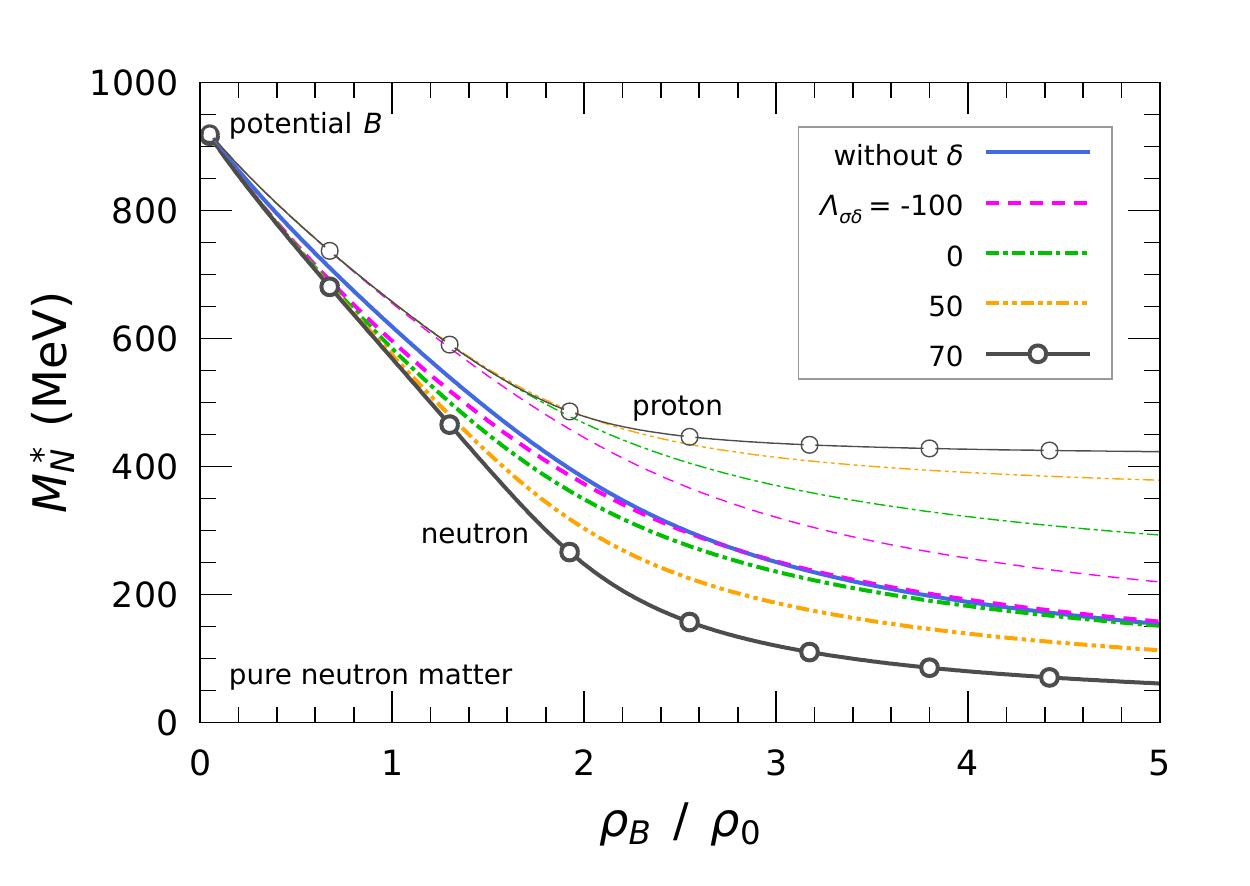}
  \caption{\label{fig:Emass-B}
    Effective nucleon mass, $M_{N=p, n}^{\ast}$, in pure neutron matter as a function of $\rho_{B}/\rho_{0}$.
    The thin (thick) lines are for proton (neutron).
    The coupling constants in potential $B$ are used.}
\end{figure}
The density dependence of effective nucleon mass, $M_{N}^{\ast}$, in pure neutron matter is expressed in Figure \ref{fig:Emass-B}.
When the $\rho$ meson only is included, the RMF model predicts the equal effective mass of proton and neutron.
However, as explained in \citet{vanDalen:2006pr}, the $\delta$ meson is responsible for the mass splitting, where the effective mass of neutron is heavier than that of proton.
It is found that, in the whole density range, the larger the coupling of $\Lambda_{\sigma\delta}$ is, the larger the mass splitting is.

In order to move on the neutron-star calculations in which the charge neutrality and $\beta$ equilibrium conditions are imposed, we introduce the degrees of freedom of leptons (electron and muon) as well as nucleons and mesons in Equation \eqref{eq:Lagrangian}.
Since the radius of a neutron star is remarkably sensitive to the nuclear EoS at very low densities, we adopt the EoS for nonuniform matter, where nuclei are taken into account using the Thomas-Fermi calculation \citep{Miyatsu:2013hea,Miyatsu:2015kwa}.

\begin{figure}
  \plotone{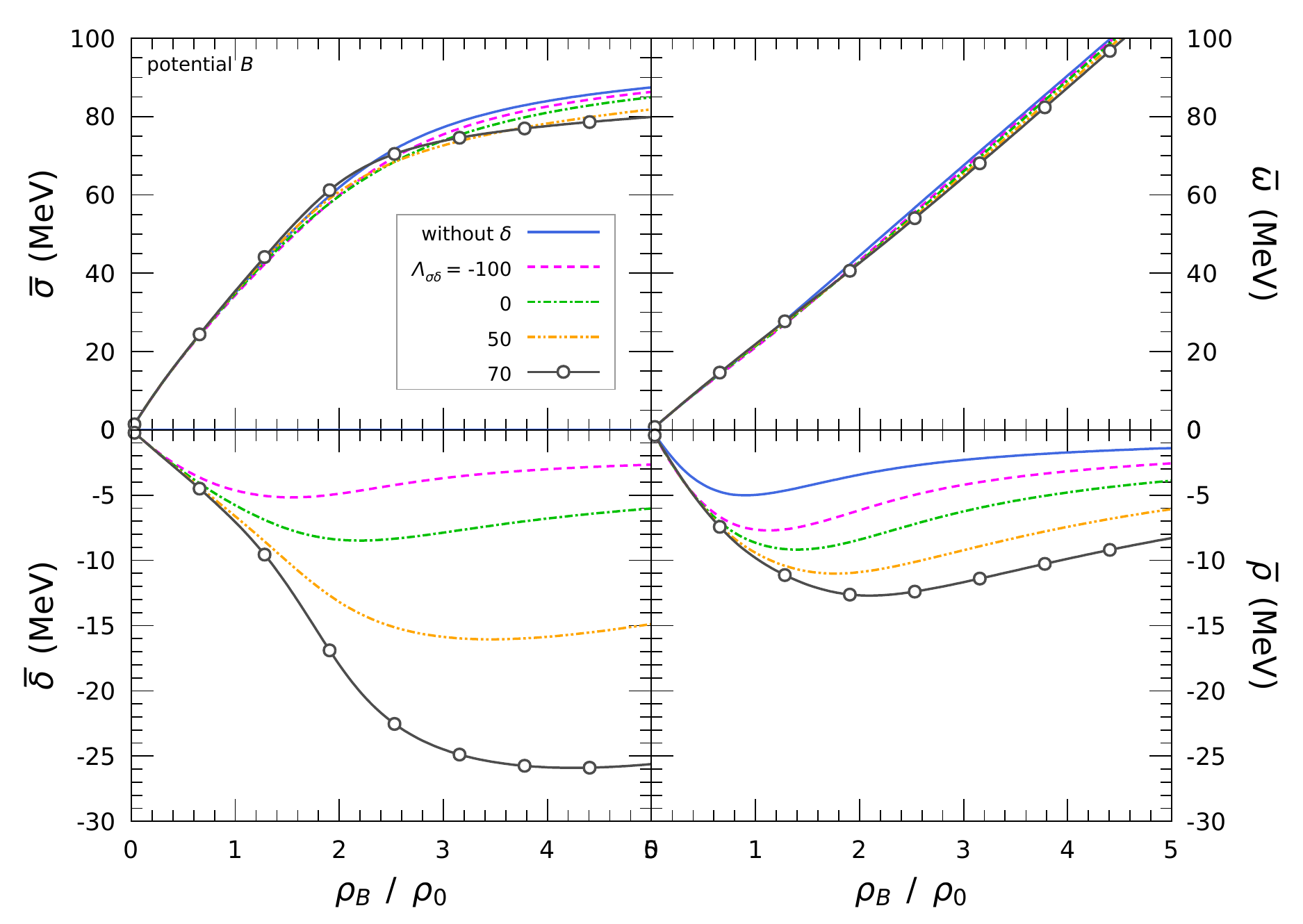}
  \caption{\label{fig:fields}
    Meson fields in neutron-star matter.
    The coupling constants in potential $B$ are used.}
\end{figure}
The meson fields in neutron-star matter are presented in Figure \ref{fig:fields}.
In the present calculation, $\Lambda_{\sigma\delta}$ has little influence on the isoscalar-meson fields, $\bar{\sigma}$ and $\bar{\omega}$, which are shown in the upper panels of Figure \ref{fig:fields}.
On the other hand, the isovector-meson fields, $\bar{\delta}$ and $\bar{\rho}$, in the lower panels of Figure \ref{fig:fields} are affected by the $\sigma$-$\delta$ mixing, namely as $\Lambda_{\sigma\delta}$ increases, $\bar{\delta}$ and $\bar{\rho}$ are (negatively) enhanced, and $\bar{\delta}$ turns out to be stronger than $\bar{\rho}$ for $\Lambda_{\sigma\delta}>0$.

\begin{figure}
  \plotone{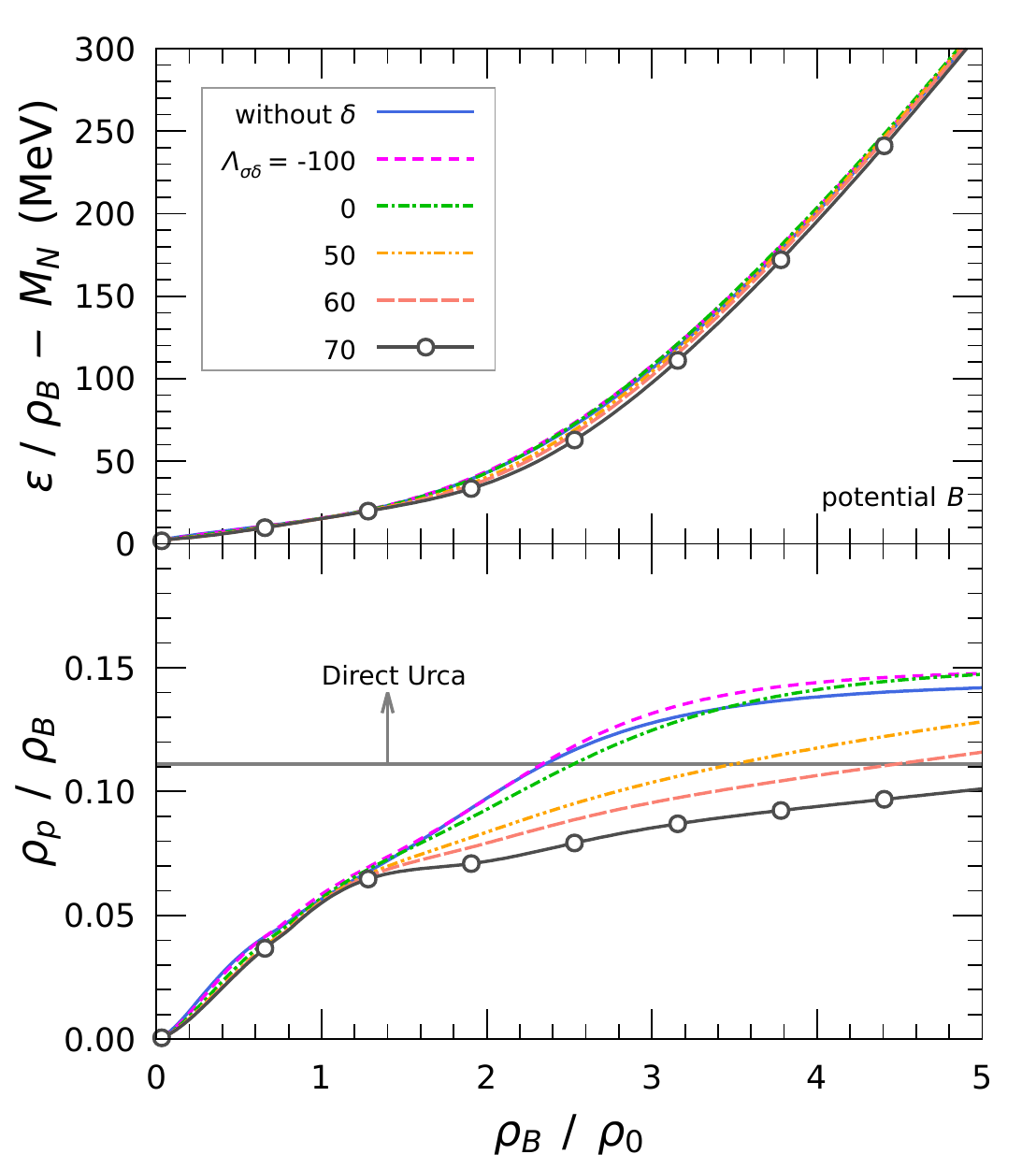}
  \caption{\label{fig:Beta}
    Binding energy per nucleon, $E_{B}$, (upper panel) and proton fraction (lower panel) in neutron-star matter.
    The coupling constants in potential $B$ are used.
    The threshold for the direct Urca process, i.e., $\rho_{p}/\rho_{B}=1/9$, is shown in the lower panel \citep{Maruyama:1999td}.}
\end{figure}
In Figure \ref{fig:Beta}, we illustrate the binding energy per nucleon, $E_{B}$, and the proton fraction in neutron-star matter.
Because the same saturation conditions are imposed in the present study, little effect due to the $\sigma$-$\delta$ mixing is seen in the binding energy, as similar to that in pure neutron matter shown in Figure \ref{fig:Bind-engy}.
One remarkable point is that the $\sigma$-$\delta$ mixing has an influence on the proton fraction at high densities.
The positive $\sigma$-$\delta$ mixing suppresses the proton fraction, and then delays the direct Urca process, in which neutrinos can be emitted rapidly.
Particularly, the direct Urca process never occurs for $\Lambda_{\sigma\delta}=70$ in the current density region, and the so-called modified Urca process, which is the standard model of neutron-star coolings, mainly takes place for the neutrino emission \citep{Lattimer:1991ib}.
On the contrary, only a small influence on the proton fraction is given for $\Lambda_{\sigma\delta}\leq0$, and the direct Urca process occurs around $2.4\rho_{0}$.

\begin{figure*}
  \plottwo{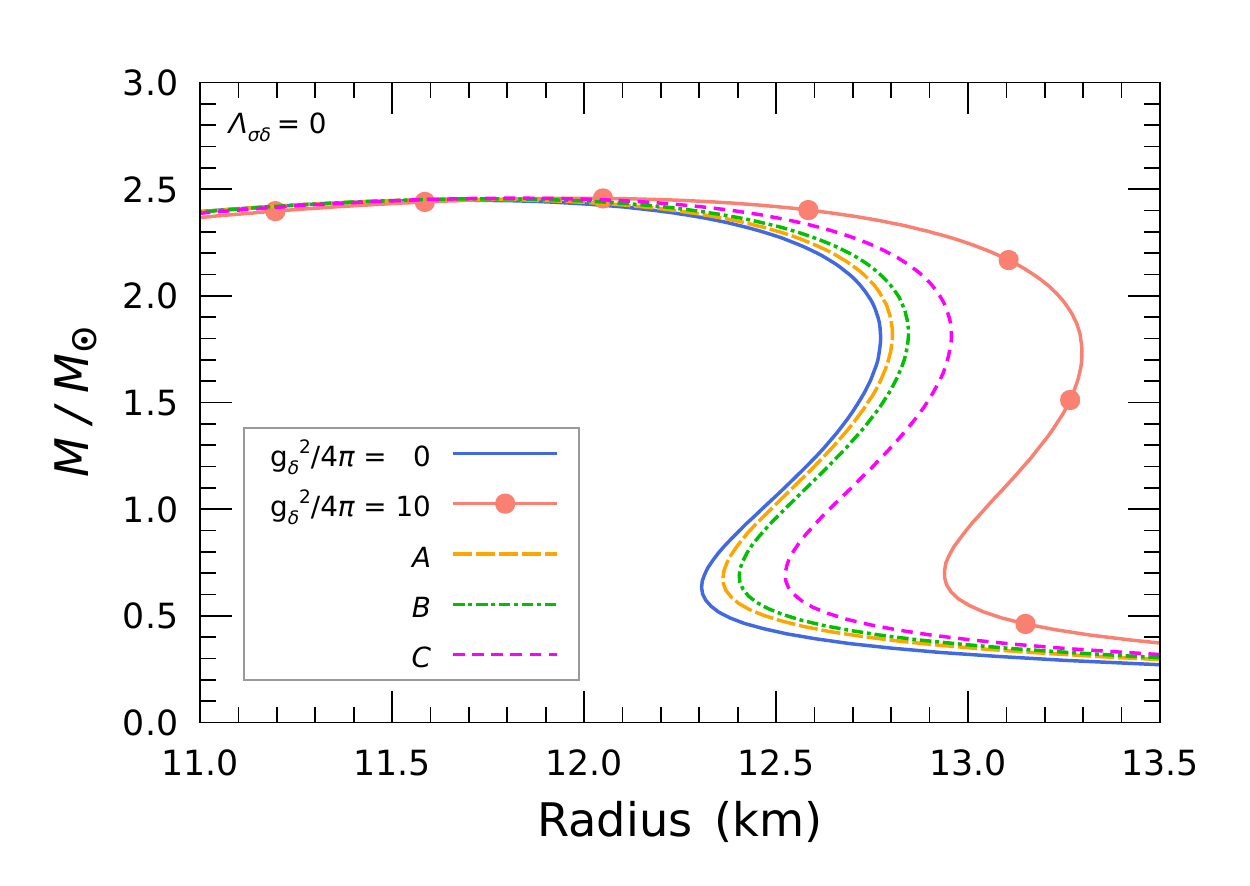}{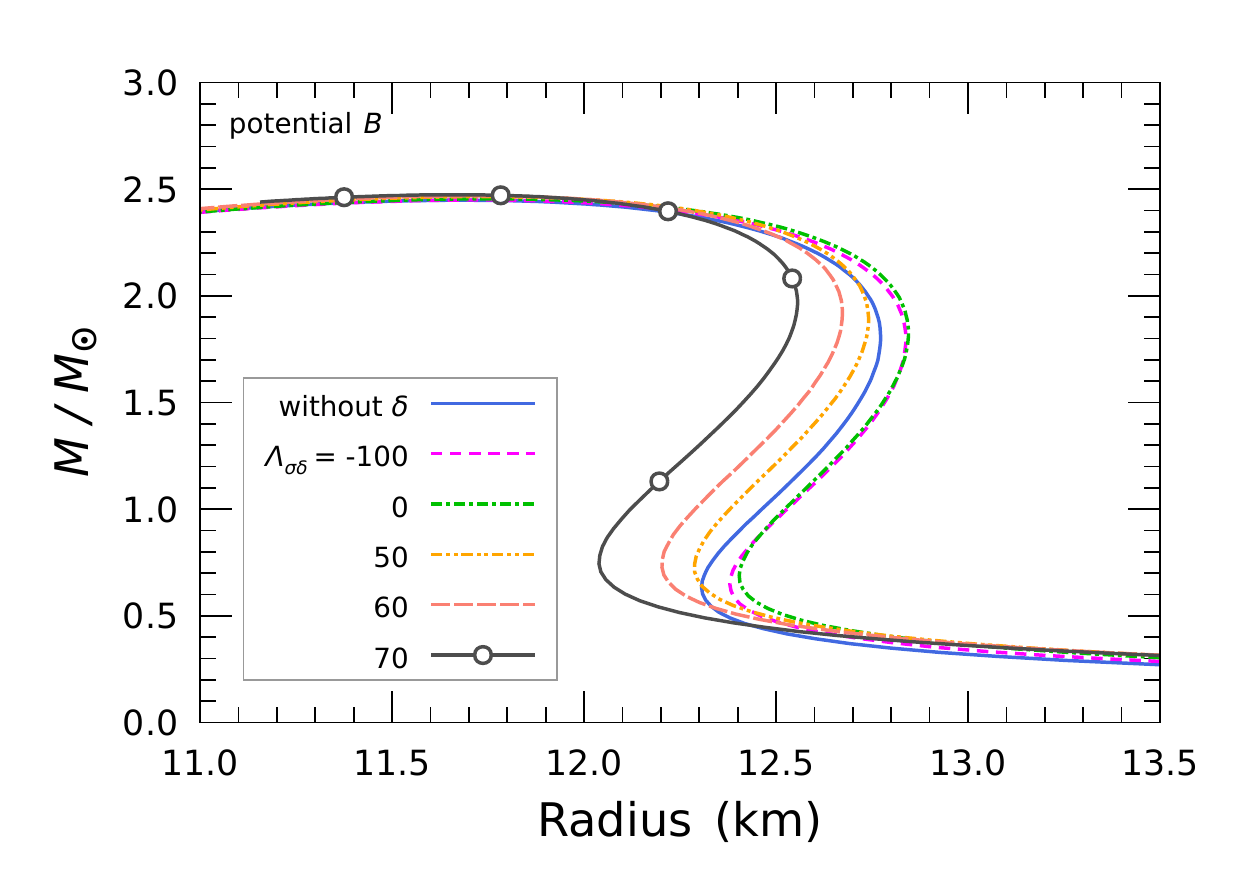}
  \caption{\label{fig:TOV}
    Mass-radius relations of neutron stars.
    The left panel is for the case in which the $\sigma$-$\delta$ mixing is ignored.
    The right panel shows the result with the $\sigma$-$\delta$ mixing in potential $B$.}
\end{figure*}
The mass-radius relations of neutron stars are presented in Figure \ref{fig:TOV}.
In both panels, the $\delta$-$N$ interaction and the $\sigma$-$\delta$ mixing have little impact on the neutron-star properties at the maximum-mass point.
Because any possibility of exotic degrees of freedom in the core is not taken into account in the present study, it is easy to support the massive neutron star, such as PSR J0740+6620 with the mass of $2.08\pm0.07M_{\odot}$ \citep{NANOGrav:2019jur,Fonseca:2021wxt}.
In all the cases, the maximum mass and its radius, $M_{\rm max}$ and $R_{\rm max}$, lie in the ranges of $2.45\le M_{\rm max}/M_{\odot}\le2.47$ and $11.66\le R_{\rm max} \mathrm{(km)} \le11.98$, respectively.
In contrast, the radius of a canonical $1.4M_{\odot}$ neutron star, $R_{1.4}$, is strongly affected by the $\delta$ meson.
The $\delta$-$N$ interaction makes $R_{1.4}$ large (see the left panel), while the $\sigma$-$\delta$ mixing reduces $R_{1.4}$ for $\Lambda_{\sigma\delta}>0$ (see the right panel).
Except for $g_{\delta}^{2}/4\pi=10$, all the results can satisfy the recent restriction, $R_{1.4}=12.45\pm0.65$ km, based on the radius measurements from {\it NICER} and {\it XMM-Newton} data \citep{Miller:2021qha}.

\begin{figure*}
  \plottwo{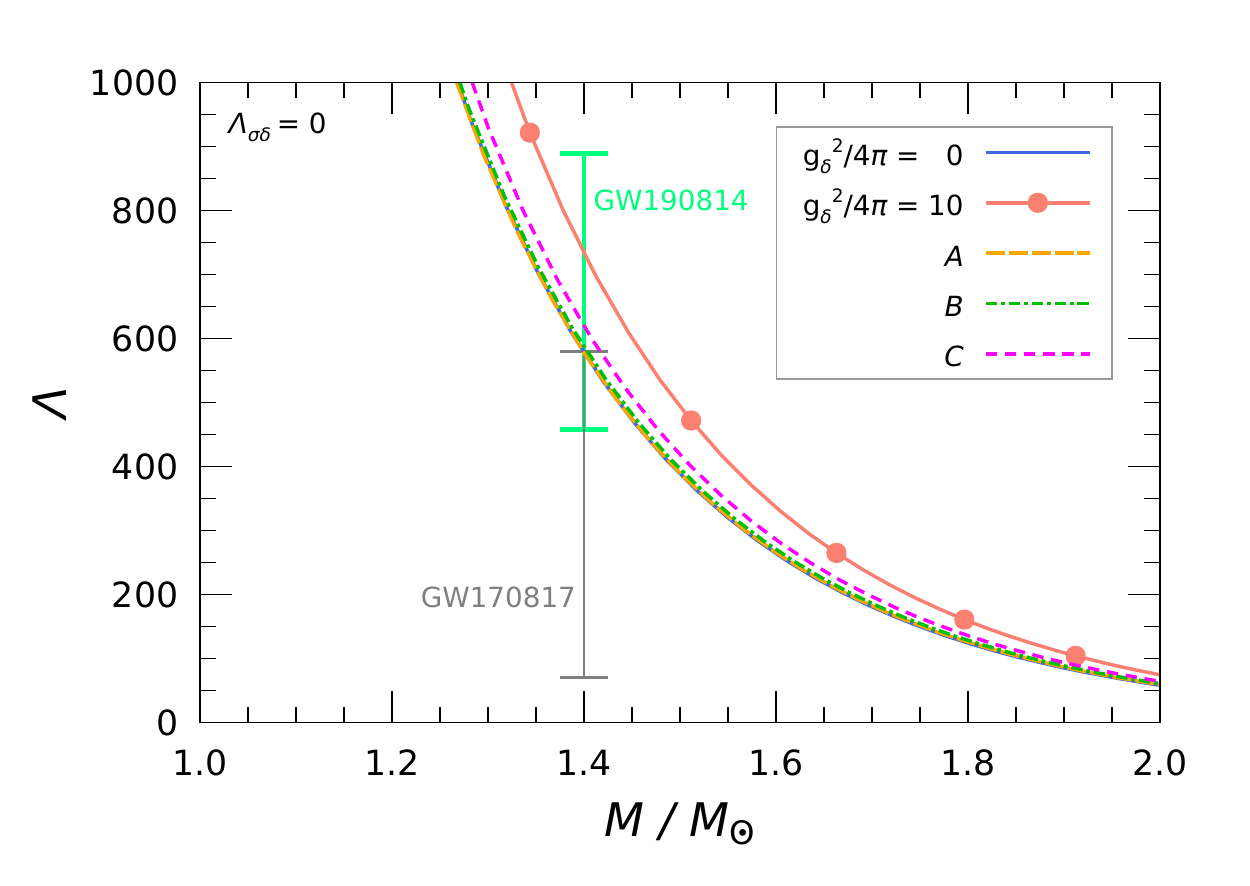}{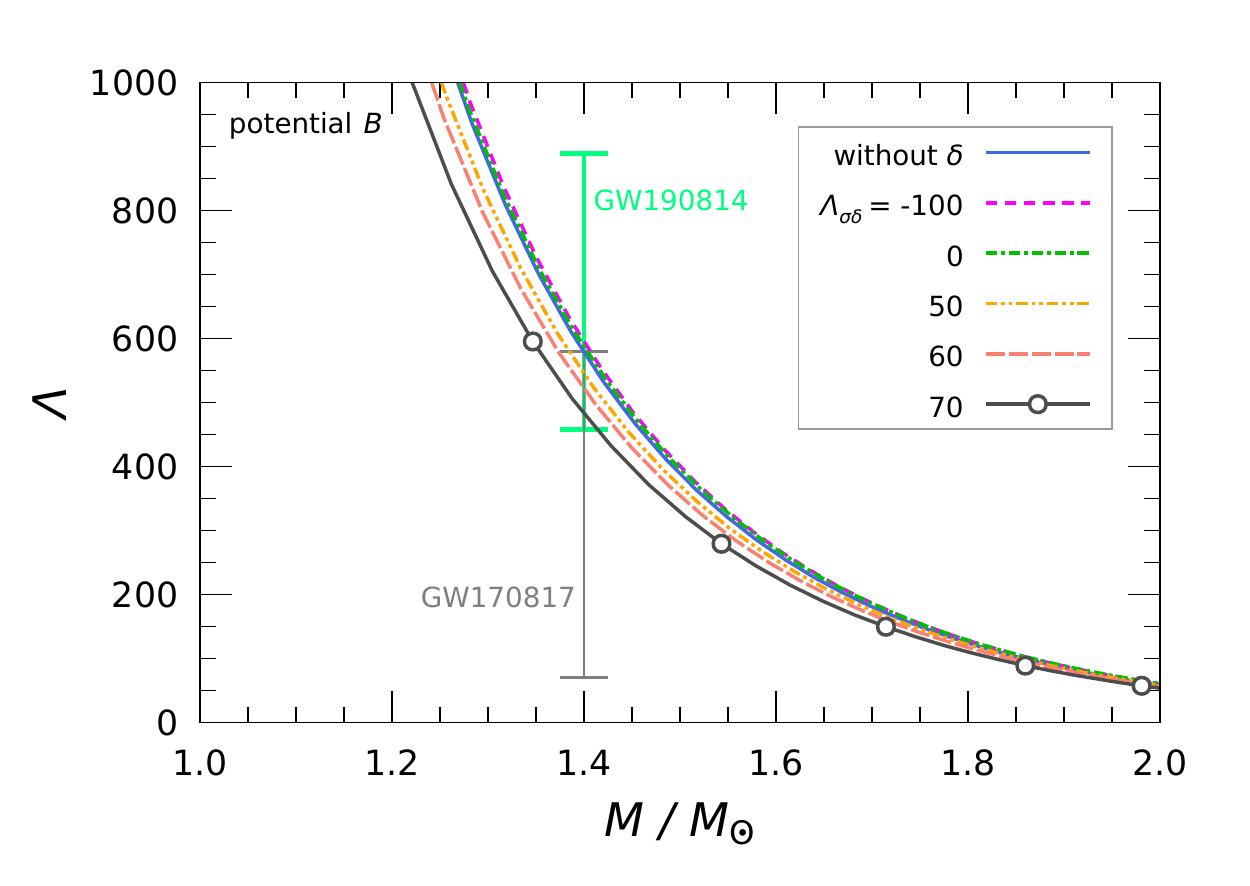}
  \caption{\label{fig:Tidal}
    Dimensionless tidal deformability of a neutron star, $\Lambda$.
    The left (right) panel is for the case without (with) the $\sigma$-$\delta$ mixing.
    In the right panel, the coupling constants in potential $B$ are used.
    The vertical bars are the astrophysical constraints on $\Lambda_{1.4}$ from binary merger events, GW170817 ($\Lambda_{1.4}=190_{-120}^{+390}$; \citet{LIGOScientific:2018cki}) and GW190814 ($\Lambda_{1.4}=616_{-158}^{+273}$; \citet{LIGOScientific:2020zkf}).}
\end{figure*}
It is quite useful to consider the tidal deformability of a neutron star because the GW signals from binary merger events can potentially yield robust information on the EoS for neutron stars \citep{Hinderer:2007mb,Hinderer:2009ca}.
In Figure \ref{fig:Tidal}, we present the dimensionless tidal deformability, $\Lambda$, which is defined as $\Lambda = \frac{2}{3}k_{2}\left(\frac{R}{M}\right)^{5}$, where $k_{2}$ is the second Love number, and $M$ and $R$ are, respectively, the mass and radius of a neutron star.
In addition, the resent constraints on tidal deformability of a canonical $1.4M_{\odot}$ neutron star, $\Lambda_{1.4}$, are given in both panels, which are based on the GW signals from the binary neutron star merger, GW170817 \citep{LIGOScientific:2018cki}, and from the compact binary coalescence involving a $23M_{\odot}$ black hole and a $2.6M_{\odot}$ compact object, GW190814 \citep{LIGOScientific:2020zkf}.

When only the $\delta$-$N$ coupling is considered, $\Lambda_{1.4}$ becomes large as $g_{\delta}^{2}$ increases (see the left panel of Figure \ref{fig:Tidal}).
It is then impossible to satisfy the astrophysical constraint on $\Lambda_{1.4}$ from GW170817, which is the highly credible data.
On the other hand, the $\sigma$-$\delta$ mixing has the promising effect on $\Lambda_{1.4}$ as seen in the right panel of Figure \ref{fig:Tidal}.
Although little impact on $\Lambda_{1.4}$ can be seen for $\Lambda_{\sigma\delta}\leq0$, the $\sigma$-$\delta$ mixing reduces $\Lambda_{1.4}$ extremely for $0<\Lambda_{\sigma\delta}\leq70$.
In particular, we can comfortably explain both constraints on $\Lambda_{1.4}$ based on the GW signals by introducing the $\sigma$-$\delta$ mixing in potential $B$.

\begin{table}[t]
  \caption{\label{tab:DATA}%
    Summary of several nuclear and neutron-star properties.}
  \begin{ruledtabular}
    \begin{tabular}{crrccc}
      \                              & $\Lambda_{\sigma\delta}$ & $K_{\rm sym}$ & $E_{\rm sym}(2\rho_{0})$ & $R_{1.4}$ & $\Lambda_{1.4}$ \\
      \                              & \                        & (MeV)         & (MeV)                    & (km)      & \               \\
      \colrule
      Without $\delta$               &                       -- &       $-60.3$ &                     49.5 &      12.7 &             579 \\
      \colrule
      \                              &                        0 &      $-114.2$ &                     49.6 &      12.7 &             578 \\
      $A$                            &                       50 &      $-178.6$ &                     45.7 &      12.6 &             540 \\
      ($g_{\delta}^{2}/4\pi=1.3$)    &                       60 &      $-197.5$ &                     43.7 &      12.5 &             515 \\
      \                              &                       70 &      $-219.3$ &                     40.7 &      12.3 &             467 \\
      \colrule
      \                              &                        0 &      $-128.4$ &                     50.6 &      12.7 &             588 \\
      $B$                            &                       50 &      $-216.6$ &                     45.3 &      12.6 &             552 \\
      ($g_{\delta}^{2}/4\pi=2.488$)  &                       60 &      $-242.1$ &                     42.5 &      12.5 &             523 \\
      \                              &                       70 &      $-271.3$ &                     38.2 &      12.4 &             483 \\
      \colrule
      \                              &                        0 &      $-123.0$ &                     54.3 &      12.9 &             620 \\
      $C$                            &                       50 &      $-231.4$ &                     49.4 &      12.8 &             614 \\
      ($g_{\delta}^{2}/4\pi=4.722$)  &                       60 &      $-262.5$ &                     46.5 &      12.8 &             608 \\
      \                              &                       70 &      $-297.8$ &                     41.9 &      12.8 &             599 \\
    \end{tabular}
  \end{ruledtabular}
\end{table}
Several properties of asymmetric nuclear matter and neutron stars, $K_{\rm sym}$, $E_{\rm sym}(2\rho_{0})$, $R_{1.4}$, and $\Lambda_{1.4}$, are listed in Table \ref{tab:DATA}.
We here discuss their dependence on the coupling strength of $g_{\delta}^{2}$ and the mixing $\Lambda_{\sigma\delta}$.
All the results of $E_{\rm sym}(2\rho_{0})$ shown in Table \ref{tab:DATA} satisfy the restriction based on nine new analyses of neutron-star observables since GW170817 ($38\leq E_{\rm sym}(2\rho_{0})\mathrm{(MeV)}\leq64$; \citet{Li:2021thg}).
Meanwhile, it is impossible for potential $C$ to explain the astrophysical constraint on $\Lambda_{1.4}$ from GW170817 ($70\leq\Lambda_{1.4}\leq580$; \citet{LIGOScientific:2018cki}).
Furthermore, if we adopt the $\sigma$-$\delta$ mixing in potential $A$ or $B$, $K_{\rm sym}$ is respectively estimated to be $-219\leq K_{\rm sym}\mathrm{(MeV)}\leq-114$ or $-271\leq K_{\rm sym}\mathrm{(MeV)}\leq-128$, which are consistent with the recent calculations by \citet{Li:2021thg}, $K_{\rm sym}=-107\pm88$ MeV, and \citet{Gil:2021ols}, $-150\leq K_{\rm sym}\mathrm{(MeV)}\leq0$, but are relatively smaller than our previous result, $-84\leq K_{\rm sym}\mathrm{(MeV)}\leq-10$ \citep{Choi:2020eun}.
As seen in Figure \ref{fig:TOV}, all the calculated $R_{1.4}$ are consistent with the observed results from {\it NICER} and {\it XMM-Newton} data \citep{Miller:2021qha}.

Finally, we give a comment on the large values of $E_{\rm sym}$ and $L$ recently deduced from PREX-II data.
Using data from two experimental runs, PREX-I and PREX-II, the PREX Collaboration has reported an unexpected thick neutron skin in $^{208}$Pb \citep{PREX:2021umo}.
It also implies the larger $L$ than any latest results obtained from other experiments and microscopic calculations \citep{Piekarewicz:2021jte,Reed:2021nqk}.
If, in the present study, we adopt the higher values, $E_{\rm sym}=38.0$ MeV and $L=100.0$ MeV, at $\rho_0$ to fix the coupling constants related to the properties of isospin-asymmetric matter, the large $\sigma$-$\delta$ mixing then makes it possible to satisfy the constraint on $E_{\rm sym}$ based on heavy-ion collision data and recent analyses of neutron-star observations as shown in Figures \ref{fig:Esym} and \ref{fig:Esym-B}.
However, even if we introduce the large $\sigma$-$\delta$ mixing, it is impossible to support the astrophysical constraints on $R_{1.4}$ based on the radius measurements from {\it NICER} and {\it XMM-Newton} data \citep{Miller:2021qha}, and $\Lambda_{1.4}$ from the neutron-star merger event, GW170817 \citep{LIGOScientific:2018cki,LIGOScientific:2018hze}.
Therefore, we emphasize that it is very hard to simultaneously explain the recent PREX-II experiment and the astrophysical observations concerning neutron stars \citep{Essick:2021ezp,Essick:2021kjb,Sahoo:2021fxw}.

\section{Summary and conclusion} \label{sec:summary}

We have studied the properties of isospin-asymmetric nuclear matter using the RMF model with nonlinear couplings between the isoscalar and isovector mesons.
Not only the isovector, Lorentz-vector ($\bm{\rho}^{\mu}$) meson but also the isovector, Lorentz-scalar ($\bm{\delta}$) meson has been taken into account as well as the isoscalar mesons ($\sigma$ and $\omega^{\mu}$).
Then, the mixing terms due to the isoscalar and isovector mesons, $\sigma^{2}\bm{\delta}^{2}$ and $\omega_{\mu}\omega^{\mu}\bm{\rho}_{\nu}\bm{\rho}^{\nu}$, have been introduced to investigate the density dependence of $E_{\rm sym}$ and the EoS for neutron stars in detail.

Firstly, it has been found that the $\delta$-$N$ interaction enhances $E_{\rm sym}$ at high densities.
Meanwhile, the quartic interaction due to the scalar mesons, $\sigma^{2}\bm{\delta}^{2}$, drastically affects $E_{\rm sym}^{\rm pot}$, in which $E_{\rm sym}^{\delta}$ and $E_{\rm sym}^{\rho}$ compete against each other.
In the region above $\rho_{0}$, the $\sigma$-$\delta$ mixing for $\Lambda_{\sigma\delta}\geq50$ decreases $E_{\rm sym}^{\rm pot}$ rapidly, and thus $E_{\rm sym}$ becomes temporarily very soft around $2\rho_{0}$.
Moreover, we have found that the $\sigma$-$\delta$ mixing is responsible for the large mass splitting between proton and neutron in isospin-asymmetric nuclear matter.

Secondly, we have presented the EoS for isospin-asymmetric nuclear matter, and investigated the properties of neutron stars.
It has been found that, under the charge neutrality and $\beta$ equilibrium conditions, the $\sigma$-$\delta$ mixing has a large impact on the isovector-meson fields, and, for $\Lambda_{\sigma\delta}>0$, the $\bar{\delta}$ field is more important than the $\bar{\rho}$ field.
Besides, the $\sigma$-$\delta$ mixing suppresses the proton fraction in the core of neutron star, and then delays the direct Urca process.
Furthermore, although the $\delta$ meson little contribute to the properties of a neutron star at the maximum-mass point, it gives a large influence on the properties of a canonical $1.4M_\odot$ neutron star.
In particular, we have found that $R_{1.4}$ and $\Lambda_{1.4}$ are largely reduced by means of the $\sigma$-$\delta$ mixing, and such tendency is favorable to satisfy the astrophysical constraints based on the radius measurements from {\it NICER} and {\it XMM-Newton} data \citep{Miller:2021qha} and the GW signals from GW170817 \citep{LIGOScientific:2018cki,LIGOScientific:2018hze}.
In conclusion, it is preferable to choose $g_\delta^2/4\pi \simeq 1.3$--$2.5$ and $0< \Lambda_{\sigma\delta}<60$ to be consistent with various constraints from the terrestrial experiments and the astrophysical  observations.

Lastly, we comment on future works.
It is urgent to study the influence of the $\sigma$-$\delta$ mixing on the neutron skin thickness of $^{208}$Pb, as in the case of the $\omega_{\mu}\omega^{\mu}\bm{\rho}_{\nu}\bm{\rho}^{\nu}$ mixing in finite nuclei \citep{Horowitz:2000xj,Horowitz:2001ya,Piekarewicz:2021jte,Reed:2021nqk}.
It is also interesting to consider how the effect of quark degrees of freedom inside a nucleon affects the characteristics of isospin-asymmetric nuclear matter including the $\sigma$-$\delta$ mixing \citep{Guichon:1987jp,Saito:1994ki,Saito:1994tq,Saito:2005rv,Nagai:2008ai}.
Furthermore, using relativistic Hartree-Fock or Dirac-Brueckner-Hartree-Fock approximation, we may understand the $\delta$-meson contribution to $E_{\rm sym}$ and $L$ in more detail \citep{Katayama:2012ge,Miyatsu:2011bc,Katayama:2013zya}.
It is also important to include hyperons in the neutron-star calculations, because the $\delta$-$N$ interaction and the quadratic mixing give a large influence on the proton fraction and the neutron-star cooling, to which hyperons also contribute \citep{Katayama:2015dga,Maruyama:2021ghf}.

\acknowledgments

This work is supported by the National Research Foundation of Korea (Grant Nos. NRF-2021R1A6A1A03043957, NRF-2020K1A3A7A09080134, and NRF-2020R1A2C3006177).

\bibliography{RMFdelta6.bib}{}
\bibliographystyle{aasjournal}

\end{document}